\documentstyle[nato,epsf,numreferences]{crckapb}
\newcommand{\noi}{\noindent}
\newcommand{\eq}{\begin{equation}}
\newcommand{\en}{\end{equation}}
\newcommand{\eqa}{\begin{eqnarray}}
\newcommand{\ena}{\end{eqnarray}}
\newcommand{\vp}{{\vec p}}
\newcommand{\vx}{{\vec x}}
\newcommand{\lra}{\longrightarrow}
\begin{opening}
\title{LORENTZ GAUGE FIXING AND LATTICE QED~$^1$}
\author{I.L. BOGOLUBSKY}
\author{V.K. MITRJUSHKIN}
\institute{
 Joint Institute for Nuclear Research,  \\
 141980 Dubna, Russia}
\author{M. M\"ULLER-PREUSSKER}
\author{P. PETER}
\author{N.V. ZVEREV}
\institute{
 Humboldt-Universit\"at zu Berlin, Institut f\"ur Physik, \\
 D-10115 Berlin, Germany}
\end{opening}
\begin{document}
\begin{abstract}
%---------------
The Gribov ambiguity problem is studied for compact lattice QED
within the Lorentz gauge. In the Coulomb phase, Gribov copies are
mainly caused by double Dirac sheets and zero-momentum modes of
the gauge fields. Removing them by (non-) periodic gauge
transformations allows to reach the absolute extremum of
the Lorentz gauge functional. For standard Lorentz gauge fixing
the Wilson fermion correlator turns out to be strongly effected by
the zero-momentum modes. A reliable fermion mass estimate requires
the proper treatment of these modes.%
\footnotetext[1]{~Talk by M. M\"uller-Preussker at the NATO Advanced
 Research Workshop  {\it Lattice Fermions and Structure of the Vacuum},
 October 1999, Dubna, Russia.}
\end{abstract}

\vspace*{-13cm}\noindent
January 15, 2000 \hfill HU-EP-00/07

\vspace*{13cm}

\section{Introduction}
%---------------------
Most of the applications of lattice gauge theories are based on and are
employing their manifest gauge invariance. However, in order
to get a better understanding of the structure of the lattice theory itself
and to interprete correctly results obtained in Monte Carlo simulations,
it is instructive to compare also gauge variant quantities such
as gauge and fermion field propagators with corresponding analytic
perturbative results. In this respect, compact lattice QED
within the Coulomb phase serves as a very useful `test ground'. In
the weak coupling limit this theory is supposed to describe
non-interacting massless photons.

In order to fix the gauge the Lorentz (or Landau) gauge condition
is normally applied. For non-Abelian gauge theories there is no unique
solution, i.e. so-called Gribov copies occur \cite{Gr}. Within
continuum QED such a problem arises, too, if the theory is defined
on a torus \cite{Ki}. The lattice discretization may cause additional
problems. Indeed, various lattice studies [3 -- 11]
%\cite{NaPl,BorMiMPPa,deFoHe,Mi1,Mi2,BogMiMPPe,BogDDMi,NaSi,BogMiMPPeZv}
have revealed nontrivial effects. The standard Lorentz
(or Landau) gauge fixing procedure leads to a
$~\tau$--dependence of the non-zero-momentum transverse photon correlator
inconsistent with the expected zero-mass behavior
\cite{NaPl}. Numerical \cite{BorMiMPPa,deFoHe,BogDDMi} and analytical
\cite{Mi2} studies have shown that there is a connection between `bad'
gauge (or Gribov) copies and the appearance of periodically closed double
Dirac sheets (DDS).  The removal of DDS by appropriate gauge transformations
restores the correct perturbative behavior of the photon correlator
at non-zero momentum, but it does not completely resolve the Gribov
ambiguity problem. Gribov copies related to other
local extrema of the gauge functional and connected with
zero--momentum modes (ZMM) of the gauge fields still appear. They
`damage' gauge dependent observables such as the zero--momentum gauge field
correlator \cite{Mi2,BogDDMi,BogMiMPPe} and the fermion propagator
\cite{NaSi,BogMiMPPeZv}, respectively.

There is a special Lorentz gauge, for which both the double Dirac sheets
and the zero-momentum modes can be removed from the gauge fields .
We call it {\it zero--momentum Lorentz gauge} (ZML) \cite{BogMiMPPe}.
It allows to reach the global extremum of the Lorentz gauge
functional in almost $100\%$ of the cases.
In comparison with the standard Lorentz gauge procedure (LG) it
demonstrates very clearly the strong effects caused by the
zero--momentum modes.

In the given talk we are going to review the results
of \cite{BogMiMPPe,BogMiMPPeZv} with special emphasis on the question,
how Gribov copies influence the Wilson--fermion propagator
within the Coulomb phase of quenched QED. We want to
show that a reliable estimate of the (renormalized)
fermion mass requires either the removal of the zero-momentum modes or
their proper perturbative treatment, when comparing the numerical
results with analytic expressions. By employing the ZML-gauge we shall
estimate the fermion mass in agreement with standard perturbation
theory.

\section{The Action and the Observables}
%---------------------------------------
We consider 4d compact QED in the quenched approximation
on a finite lattice ($V=N_s^3 \times N_t$).
The pure gauge part of the standard Wilson action \cite{Wi} reads
\eq \label{1}
S_G = \beta\sum_{x,\mu < \nu}\left(1-\cos\theta_{x,\mu\nu}\right),
\en
with the plaquette angle
$\theta_{x,\mu\nu}=\theta_{x,\mu}+\theta_{x+\hat{\mu},
\nu}-\theta_{x+\hat{\nu},\mu}-\theta_{x,\nu}$
related to the link variables
$\theta_{x,\mu}\in (-\pi, \pi].$
$\beta=1/e_0^2$ is the inverse bare coupling.
The lattice spacing is put $a=1$, for simplicity.

\noi The fermion part  is given by
\eq \label{2}
S_F=\sum_{x,y}\overline{\psi}_{x}{\bf M}_{xy}(\theta)\psi_{y},
\qquad {\bf M}={\bf 1} - \kappa {\bf D},
\en
$$
{\bf D}_{xy}=\sum_{\mu=1}^{4}
    \Bigl\{U_{x,\mu}P_{\mu}^{-}\delta_{y,x+\hat{\mu}}
          + U^{*}_{x-\hat{\mu},\mu} P_{\mu}^{+}\delta_{y,x-\hat{\mu}}\Bigr\},
$$
where $P_{\mu}^{\pm} = \hat{1} \pm \gamma_\mu$ and
$U_{x,\mu} = {\rm e}^{ {\rm i} \theta_{x,\mu}}.$
The hopping-parameter $\kappa$ is related to the bare mass $~m_0~$
by $~\kappa = 1/(8+2 m_0).$

In quenched QED the observables have to be averaged with respect
to the gauge field $~\{\theta\}~$ with the weight $~\exp(-S_G)~$.
We imply periodic boundary conditions (b.c.) except for the fermion
fields, which we choose to be anti-periodic in the 
$x_4 \equiv \tau$-direction.

The first gauge variant observable we are going to discuss is the
transverse photon correlator at non-zero momentum
\eqa
 \Gamma^{\rm ph}_T(\vp;\tau)~=~
            \langle \Phi(\vp;\tau) ~\Phi^{\ast}(\vp;0) \rangle ~, \\
 \Phi(\vp;\tau)~=~
 \sum_{\vx} \exp(i \vp \vx +\frac{i}{2}p_{\mu}) ~\sin \theta_{\vx \tau, \mu}
 \nonumber
\ena
with $(~\mu=1,3, ~~\vp=(0,p,0)~).$
The second one is the fermion propagator.  For a given
gauge field $~\{\theta\}~$ we have
\eq \label{4a}
\Gamma(\tau)=\frac{1}{V}\sum_{\vec{x},x_4}\sum_{\vec{y}}
{\bf M}^{-1}_{\vec{x},x_4; \vec{y},x_4 + \tau}(\theta).
\en
In the following we shall restrict ourselves to the vectorial part
\eq \label{4b}
\Gamma_V(\tau) =\frac{1}{4}{\rm Re\,Tr\,}
\left(\gamma_4\Gamma(\tau)\right),
\en
with the trace taken with respect to the spinor indices.
For the b.c. mentioned above, $\langle\Gamma_V(\tau)\rangle$ 
is an even function of 
$\tau - N_t/2.$

Lateron, we shall compare the expectation value
$~\langle~\Gamma_V~\rangle~$ with the result of a simple
approximation, which takes only constant gauge field modes
into account. The correlator in a uniform background
$\theta_{x,\mu}\equiv\phi_{\mu},~-\pi<\phi_{\mu}\le\pi,$ $\mu=1,\cdots,4~$
can be represented as
\eqa  \label{6b}
\Gamma_V(\tau;\phi) &=& \frac{1-\delta_{\tau,0}} {2(1+{\cal M})}
\times \\
&\times& \frac{[{\cal E}^{\tau}+{\cal E}^{2N_t-\tau}]\cos(\phi_4\tau)
+[{\cal E}^{N_t+\tau}+{\cal E}^{N_t-\tau}]\cos[\phi_4(N_t-\tau)]}
{1+{\cal E}^{2N_t}+2{\cal E}^{N_t}\cos(\phi_4 N_t)},
\nonumber
\ena
where
$$
{\cal E}= 1 + \frac{{\cal M}^2+{\cal K}^2}{2(1+{\cal M})}
- \frac{\sqrt{{\cal M}^2+{\cal K}^2}\sqrt{({\cal M}+2)^2+{\cal K}^2}}
{2(1+{\cal M})} ~;
$$
$$
{\cal M}=m_0+\sum_{l=1}^{3}\left(1-\cos \phi_l\right), \qquad
{\cal K}=\sqrt{\ \sum_{l=1}^{3}\sin^2 \phi_l}, \qquad m_0>0.
$$
For $~\phi_{\mu}=0, ~~\mu=1,\cdots,4~$
the free fermion correlator for finite lattice size \cite{CaBa}
is reproduced.

\section{Lorentz Gauge Fixing}
%-----------------------------
In numerical simulations the Lorentz gauge is fixed by
iteratively maximizing the gauge functional
\eq
F(\theta) = \frac{1}{V_4} \sum_{x} F_{x}(\theta )~;
\qquad F_{x}(\theta) = \frac{1}{8}\sum_{\mu=1}^4
\Bigl[ \cos \theta_{x\mu}+ \cos \theta_{x-{\hat \mu};\mu}\Bigr]
                                                 \label{10}
\en
\noi with respect to the (local) gauge transformations
\eq
U_{x\mu} \lra \Lambda_{x} U_{x \mu} \Lambda_{x+{\hat \mu}}^{\ast}~;
~~\Lambda_x=\exp\{i\Omega_x\} \in U(1)~.
                                                 \label{10a}
\en
The algorithm is called standard Lorentz gauge fixing (LG), if it
consists only of local maximization and overrelaxation steps \cite{MaOg}
with respect to gauge transformations periodic in space-time.
The standard procedure gets normally stuck into {\it local}
maxima of the gauge functional (\ref{10}) (gauge copies).
%%%%%%%%% Figure 1 %%%%%%%%
\begin{figure}[pt]
\vspace{0.5cm}
\begin{center}
\leavevmode
\hbox{
\epsfysize=8cm
\epsfxsize=8cm
\epsfbox{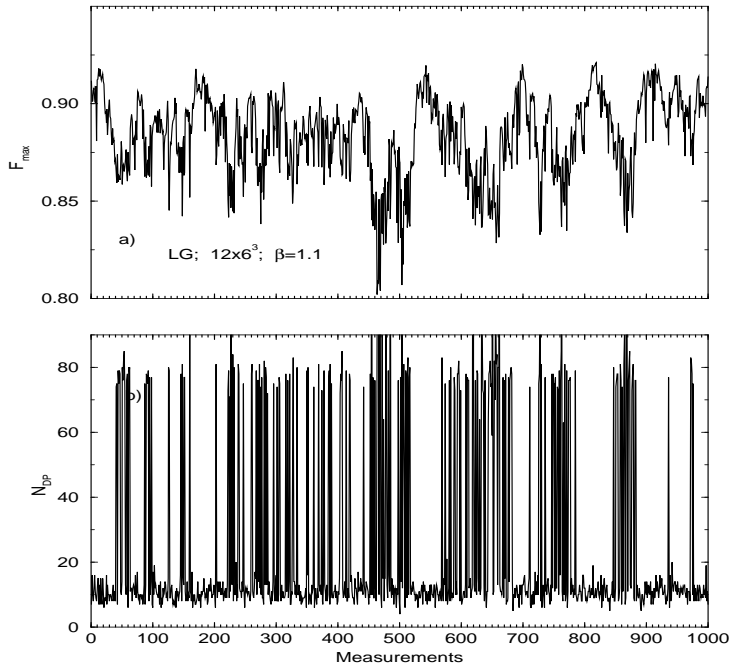}
}
\end{center}
\vspace{0.2cm}
\caption{
Time history of $F_{max}$ ({\bf a}) and $~N_{DP}$
({\bf b}) at $\beta=1.1$ on the $12\cdot 6^3$ lattice in
the standard Lorentz gauge.
}
\label{fig:fmax_12x06_b01p10_LG}
\end{figure}
%%%%%%%%%%%%%%%%%%%%%%%%%%%
%%%%%%%%% Figure 2 %%%%%%%%
\begin{figure}[pt]
%\vspace{0.5cm}
\begin{center}
\leavevmode
\hbox{
\epsfysize=9cm
\epsfxsize=9cm
\epsfbox{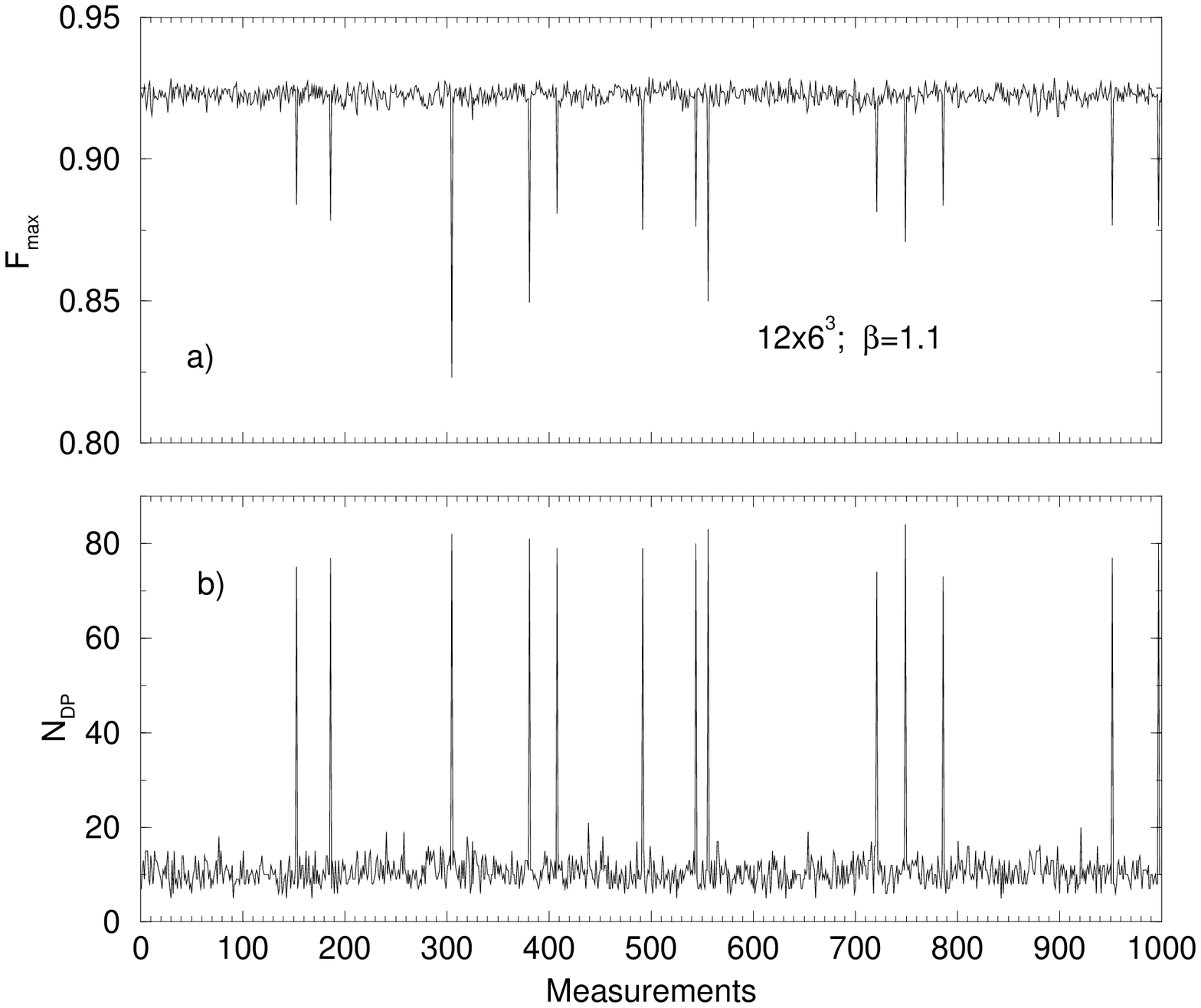}
}
\end{center}
%\vspace{0.5cm}
\caption{
Time history of $F_{max}$ ({\bf a}) and $~N_{DP}$
({\bf b}) at $\beta=1.1$ on the $12\cdot 6^3$ lattice in
ZML gauge.
}
\label{fig:fmax_12x06_b01p10_ZML}
\end{figure}
%%%%%%%%%%%%%%%%%%%%%%%%%%%
It has been argued that the Gribov problem has to be solved by searching
for the {\it global maximum} providing the best gauge copy  \cite{Zw}.
In \cite{BogMiMPPe} we have shown that in order to reach the global maximum
we have necessarily to suppress both the double Dirac sheets (DDS) and the
zero-momentum modes (ZMM) in the gauge fields. Let us explain this in more 
detail.

DDS can be identified as follows. The plaquette angle $~\theta_{x,\mu\nu}~$
is decomposed into the gauge invariant (electro-) magnetic flux
$~\overline{\theta}_{x,\mu\nu} \in (-\pi,\pi]~$
and the discrete gauge-dependent contribution
$~2 \pi n_{x,\mu\nu}, ~~n_{x,\mu\nu}=0,\pm 1,\pm 2$ \cite{DeGrTo}.
The latter represents a Dirac string passing through the given plaquette if
$~n_{x,\mu\nu}=\pm 1~$ ({\it Dirac plaquette}). A set of Dirac
plaquettes providing a world sheet of a Dirac string on the dual
lattice is called {\it Dirac sheet}.
DDS consist of two sheets with opposite flux orientation
extending over the whole lattice and closing themselves
by the periodic b.c.
They can easily be identified by counting the total number of
Dirac pla\-quettes $N^{(\mu\nu)}_{DP}$ for each choice $(\mu;\nu)$.
The necessary condition for the occurence of DDS is that at least
for one of the six possibilities $(\mu;\nu)$ holds
\eq  \label{12}
N^{(\mu\nu)}_{DP} \ge 2\frac{V}{N_{\mu}N_{\nu}}.
\en
DDS can be removed by periodic gauge transformations.

The ZMM of the gauge field
\eq \label{13}
\phi_{\mu}=\frac{1}{V}\sum_{x}\theta_{x,\mu}
\en
do not contribute to the pure gauge field action either.
For gauge configurations representing small fluctuations around
constant modes it is easy to see, that the global maximum of the functional
(\ref{10}) requires $~\phi_{\mu}\equiv 0~$.
The latter condition can be achieved by non-periodic gauge transformations
\eq
\theta_{x,\mu} \rightarrow \theta^{\ c}_{x,\mu}=c_{\mu}+\theta_{x,\mu}
\quad {\rm mod\ }2\pi, \quad c_{\mu} \in (-\pi,\ \pi].
\label{15}
\en
We realize a proper gauge fixing procedure as proposed
in \cite{BogMiMPPe}. Successive Lorentz gauge iteration steps are always
followed by non-periodic gauge transformations suppressing the ZMM.
Additionally we check, whether the gauge fields contain
yet DDS. The latter can be excluded by repeating the procedure with
initial random gauges.
We call the combined procedure {\it zero-momentum Lorentz gauge} (ZML gauge).
It yields the global maximum of the gauge functional with very high accuracy.

In Figures \ref{fig:fmax_12x06_b01p10_LG} and
\ref{fig:fmax_12x06_b01p10_ZML}
we show, how the achieved values of the gauge functional (\ref{10})
are correlated with the occurence of DDS visible as sharp peaks
in the number of Dirac plaquettes. Whereas for LG strong fluctuations
occur, they disappear after ZML gauge. The few
DDS seen in Fig. \ref{fig:fmax_12x06_b01p10_ZML}
are easily removed by restarting the procedure
with random initial gauges. Random gauges can also be used
in order to convince oneself that the ZML gauge prescription
leads to the global maximum of the gauge functional in more
than $99\%$ of the cases.

\section{Results}
%----------------
First let us convince ourselves that the removal of the above mentioned
gauge copies leads to the correct behaviour of the transverse photon
propagator.
%%%%%%%%% Figure 3 %%%%%%%%
\begin{figure}[pt]
%\vspace{0.5cm}
\begin{center}
\leavevmode
\hbox{
\epsfysize=7cm
\epsfxsize=9cm
\epsfbox{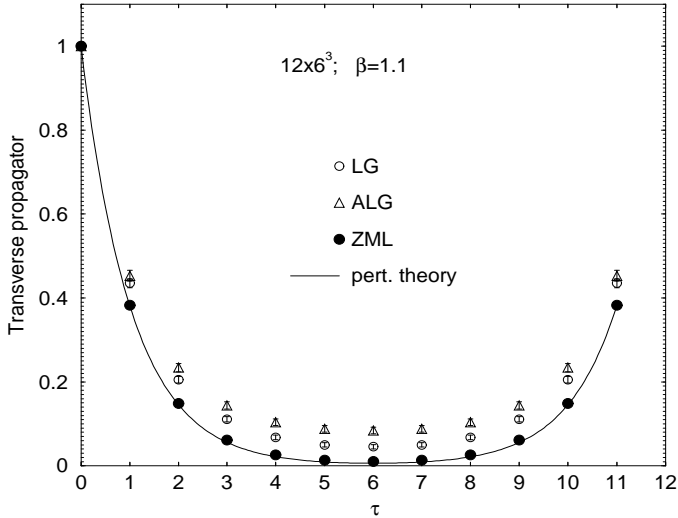}
}
\end{center}
%\vspace{0.3cm}
\caption{
Transverse propagator at $\beta=1.1$ on the $12\cdot 6^3$ lattice
in three different gauges.
}
\label{fig:pcr_12x06_b01p10_3gauge}
\end{figure}
%%%%%%%%%%%%%%%%%%%%%%%%%%%%%%%%%%%%%%%%%%
In Fig. \ref{fig:pcr_12x06_b01p10_3gauge}
we show the normalized correlator
$~\Gamma^{ph}_T(\vp;\tau) / \Gamma^{ph}_T(\vp;0)~$
for lowest non-vanishing momentum and for different Lorentz gauge
prescriptions. For the standard one (LG) we see a clear deviation
from the expected perturbative zero-mass result. We show also the result
obtained with an axial Lorentz gauge (ALG) using an initial maximal-tree
axial gauge condition \cite{Cr}, which provides a 'unique' prescription.
The latter gauge fixing prescription turns out to be even worse! On the other
hand the ZML gauge provides an excellent agreement with the perturbative
result. In fact, as we convinced ourselves earlier, it is already sufficient
to remove the DDS gauge copies in order to reach this agreement
\cite{BorMiMPPa}. The given observations do not change, when
$~\beta~$ and/or the lattice size are increased considerably
\cite{BogDDMi}.

In the following we want to concentrate on the pure effect of the ZMM.
Therefore, we compare the ZML gauge with a version of the standard
Lorentz gauge, where the DDS are removed and the ZMM are left. We shall
abbreviate the latter version also by LG.
%%%%%%%%%  Figure 4  %%%%%%%%%%%%%%%%
\begin{figure}[tbp]
%\vspace{0.5cm}
\begin{center}
\leavevmode
\hbox{
\epsfysize=10cm
\epsfxsize=8cm
\epsfbox{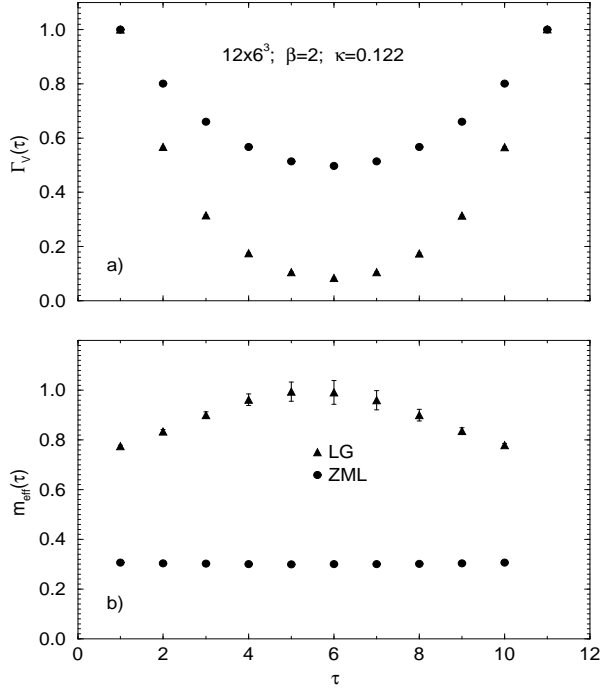}
}
\end{center}
\vspace{-0.5cm}
\caption{The fermion vector propagator ({\bf a}) and the effective
mass ({\bf b}) at $\beta=2$ and $\kappa=.122$ on a
$12\times 6^3$ lattice for LG and ZML gauges as explained in the text.}
\label{fig:fc_meff_12x06_b02p00_k122_2gau}
\end{figure}
%%%%%%%%%%%%%%%%%%%%%%%%%%%%%%%%%%%%
For both the LG and ZML gauges we have computed the averaged fermion correlator
employing the conjugate gradient method and point-like sources.
In the upper part of Fig. \ref{fig:fc_meff_12x06_b02p00_k122_2gau} we have
plotted $~\Gamma_V(\tau)~$
(normalized to unity at $\tau=1$). The situation seen is typical for
a wide range of parameter values within the Coulomb phase. Obviously,
there is a strong dependence of the fermion propagator
on the gauge fixing procedure resulting in the presence or absence of ZMM.
The masses to be extracted seem to have different values.
Let us determine the effective mass $~m_{eff}(\tau)$ in accordance with
\eq \label{16a}
\frac{\langle\Gamma(\tau+1;\theta)\rangle_{\theta}}
     {\langle\Gamma(\tau;  \theta)\rangle_{\theta}}
 =\frac{\cosh[E(\tau)(N_t/2-\tau-1)]}
       {\cosh[E(\tau)(N_t/2-\tau  )]}
\en
where $E(\tau)=\ln (m_{eff}(\tau)+1)$.
See the lower part of Fig. \ref{fig:fc_meff_12x06_b02p00_k122_2gau}.
In the LG case no plateau is visible, whereas
the ZML case provides a very stable one.
Thus, only the ZML gauge yields
a reliable mass estimate, whereas the standard method
to fix the Lorentz gauge obviously fails.

To get deeper insight into the effect of ZMM for the LG case
(with DDS suppressed) we measure the probability distributions $~P(\phi)~$
for the space- and time-like components of ZMM according to Eq. (\ref{13}).
The distributions turn out to be flat up to an effective
cutoff at $~|\phi_{\mu}| \simeq \pi/N_{\mu}$ and to be widely
independent of $\beta$. In accordance with Eq. (\ref{6b})
we compute the fermion propagator for constant
modes in the LG case and average
\eq
\langle\Gamma_V(\tau;\phi)\rangle_{\phi} =
\int [{\rm d} \phi]~ P(\phi)~ \Gamma_V(\tau;\phi).
\en
%%%%%%%%%%%%%  Figure 5  %%%%%%%%%%%%%%%%%%%%%%
\begin{figure}[tbp]
%\vspace{0.5cm}
\begin{center}
\leavevmode
\hbox{
\epsfysize=10cm
\epsfxsize=10cm
\epsfbox{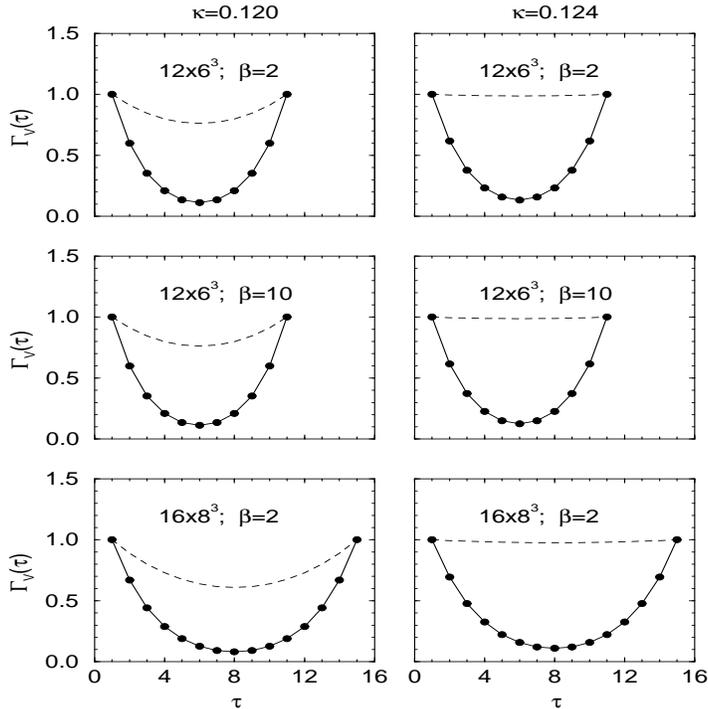}
}
\end{center}
\vspace{-0.5cm}
\caption{Free fermion propagator (dashed line)
and averaged constant-mode propagator in the LG case (full line)
for $\beta=2, 10$, $\kappa=.120, .124$, lattice size $12 \times 6^3$,
$16 \times 8^3$.
}
\label{fig:fc_tree_2lat_2bet}
\end{figure}
%%%%%%%%%%%%%%%%%%%%%%%%%%%%%%%%%%%%%%%%%%%%%
%%%%%%%%%%%%%%% Figure 6 %%%%%%%%%%%%%%%%%%
\begin{figure}[pt]
%\vspace{0.5cm}
\begin{center}
\leavevmode
\hbox{
\epsfysize=8cm
\epsfxsize=8cm
\epsfbox{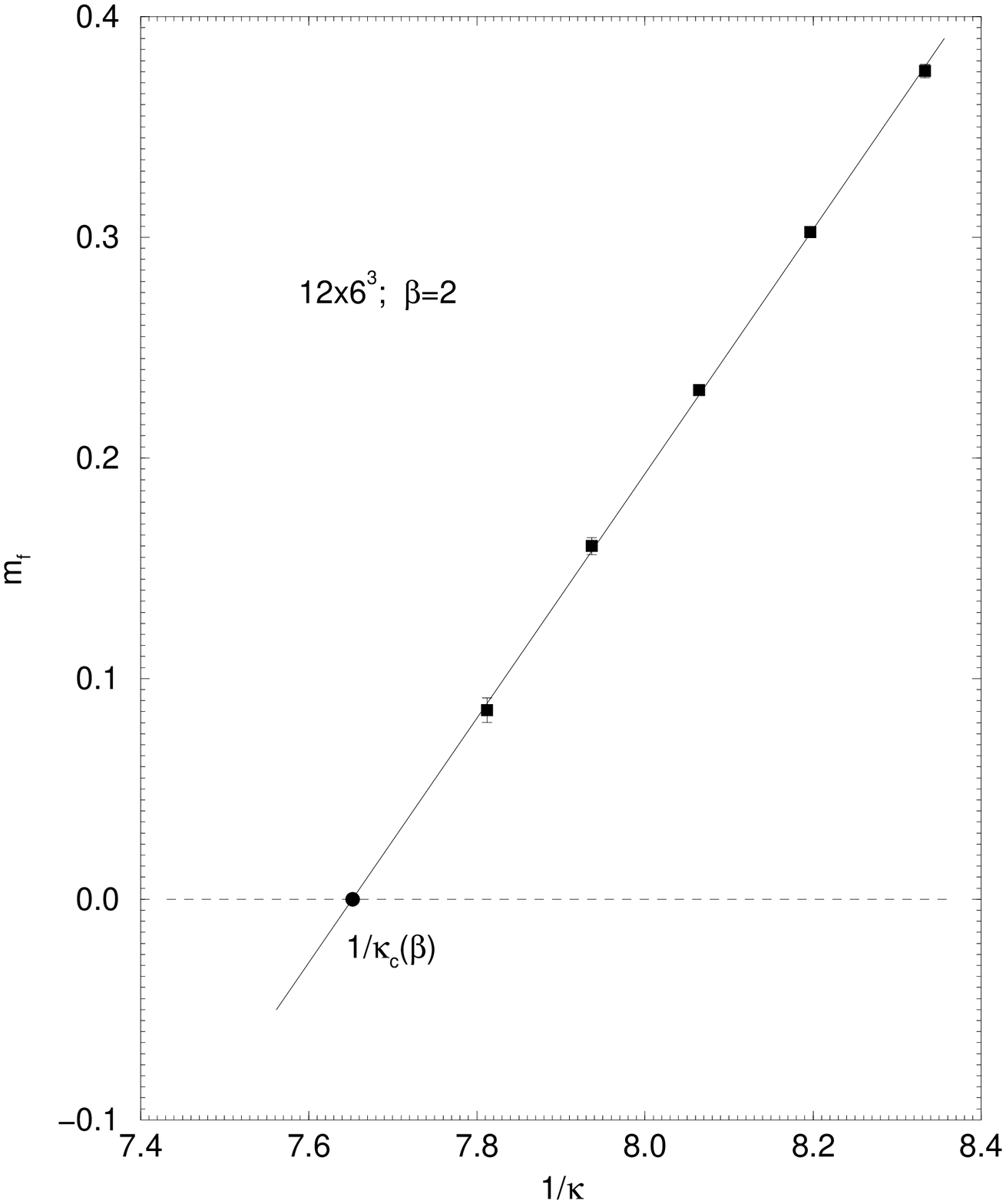}
}
\end{center}
\vspace{-0.5cm}
\caption{Fermion mass as a function of inverse $\kappa$ obtained within
the ZML gauge for $\beta=2.0$ on a $12 \times 6^3$ lattice.
The solid line represents a linear fit providing
$~\kappa_c(\beta) = 0.1307 \pm 0.0001~$.
}
\label{fig:mf_fpr}
\end{figure}
%%%%%%%%%%%%%%%%%%%%%%%%%%%%%%%%%%%%%%%%%%%%%%%%%%%
The results for several parameter sets are presented
in Fig. \ref{fig:fc_tree_2lat_2bet} together with the corresponding
free (i.e. zero-background) propagator. We see clearly that the constant
mode contributions strongly change the behavior of the fermion
propagator and, naively speaking, produce a larger mass.

Finally, in Fig. \ref{fig:mf_fpr} we present the fermion mass extracted
from the vector fermion propagator within the ZML gauge for
$\beta=2.0$ and various $\kappa$-values.
We see a nice linear behaviour
from which by extrapolating to zero mass (solid line) we estimate
the critical value $\kappa_c= 0.1307 \pm 0.0001$.

\section{Conclusions}
%--------------------
We have studied the effect of different gauge copies of the gauge
field on gauge dependent correlators, in particular on the Wilson fermion
propagator.
We have convinced ourselves that the standard Lorentz gauge fixing
prescription to maximize the functional (\ref{10}) provides gauge
copies with DDS and ZMM. These modes disturb the photon and the
fermion correlator in comparison with  perturbation theory
and consequently spoil the (effective) mass estimate.
A Lorentz gauge employing non-periodic gauge transformations in order to
suppress the ZMM -- additionally to DDS --
(the ZML gauge) allows to reach the global maximum of the Lorentz gauge
functional. Furthermore, it  provides a reliable fermion mass determination,
at least, if $\kappa$ is chosen not too close to the chiral critical
line $\kappa_c(\beta)$. A
computation of the fermion propagator with constant background
gauge fields taken from the ZMM of the quantum fields
demonstrates the disturbing effect of these modes very clearly. Moreover,
it shows the effect to be independent of the bare coupling
and not to disappear for large volumes.

So far, we have studied the quenched approximation of U(1) lattice gauge theory.
The gauge action (\ref{1}) is invariant under non-periodic gauge
transformations (\ref{15}). Thus, we are allowed to use the ZML gauge
for evaluating gauge dependent objects.
Contrary to the gauge action, the fermionic part (\ref{2}) does depend
on the ZMM because of the (anti-) periodic boundary conditions.
In this case another way of dealing with the Gribov problem
has to be searched for.

The problems we have discussed here for compact QED show that gauge fixing
has to be carried out and to be interpreted with care. This lesson has
to be taken into account also in lattice QCD when
extracting masses from gauge variant gauge and fermion correlators,
respectively.

\section*{Acknowledgements} 
%--------------------------
The work has been supported by the grant INTAS-96-370,
the RFRB grant 99-01-01230 and the JINR Dubna Heisenberg-Landau program.

\vspace{0.5cm}


\begin{thebibliography}{99}
\newcommand{\prd}[1]{{\it Phys. Rev.}, {\bf D#1}}
\newcommand{\plb}[1]{{\it Phys. Lett.}, {\bf #1B}}
\newcommand{\npb}[1]{{\it Nucl. Phys.}, {\bf B#1}}
\newcommand{\npbps}[1]{{\it Nucl. Phys. B (Proc. Suppl.)}, {\bf #1}}
%
\bibitem{Gr}        V.N. Gribov (1978), \npb{139}, p. 1.
\bibitem{Ki}        T.P. Killingback (1984), \plb{138}, p. 87.
\bibitem{NaPl}      A. Nakamura and M. Plewnia (1991),
                    \plb{255}, p. 274.
\bibitem{BorMiMPPa} V.G. Bornyakov, V.K. Mitrjushkin, M. M\"uller-Preussker
                    and F. Pahl (1993), \plb{317}, p. 596.
\bibitem{deFoHe}    Ph. de Forcrand and J.E. Hetrick (1995), \npbps{42}, p. 861.
\bibitem{Mi1}       V.K. Mitrjushkin (1996), \plb{389}, p. 713.
\bibitem{Mi2}       V.K. Mitrjushkin (1997), \plb{390}, p. 293.
\bibitem{BogMiMPPe} I.L. Bogolubsky, V.K. Mitrjushkin, M. M\"uller-Preussker
                    and P. Peter (1999), \plb{458}, p. 102.
\bibitem{BogDDMi}   I.L. Bogolubsky, L. Del Debbio and V.K. Mitrjushkin (1999),
                    \plb{463}, p. 109.
\bibitem{NaSi}      A. Nakamura and R. Sinclair (1990), \plb{243}, p. 396.
\bibitem{BogMiMPPeZv} I.L. Bogolubsky, V.K. Mitrjushkin, M. M\"uller-Preussker,
                    P. Peter and N. Zverev (1999),
                    JINR E2-99-288 and HUB-EP-99-51, hep-lat/9912017.
\bibitem{Wi}        K. Wilson (1974), \prd{10}, p. 2445.
\bibitem{CaBa}      D.B. Carpenter and C.F. Baillie (1985),
                    \npb{260}, p. 103.
\bibitem{MaOg}     J.E. Mandula and M. Ogilvie (1990), \plb{248}, p. 156.
\bibitem{Zw}        D. Zwanziger (1991), \npb{364}, p. 127;
                    (1992), {\bf B378}, p. 525.
\bibitem{DeGrTo}    T.A. DeGrand and D. Toussaint (1980), \prd{22}, p. 2478.
\bibitem{Cr}      M. Creutz (1977), \prd{15}, p. 1128.
\end{thebibliography}
\end{document}